\documentclass[a4paper]{jpconf}
\usepackage{amsmath,amssymb}
\usepackage{graphicx}

\usepackage{Braket,cite}
\usepackage{bm}

\begin{document}
\title{Trembling electrons cause conductance fluctuation}

\author{$^{*}$Yu Iwasaki, Yoshiaki Hashimoto, Taketomo Nakamura, Shingo Katsumoto}

\address{Institute for Solid State Physics, University of Tokyo, 5-1-5 Kashiwanoha, Kashiwa, Chiba 277-8581, Japan}

\ead{you.iwasaki@issp.u-tokyo.ac.jp}

\begin{abstract}
The highly successful Dirac equation can predict peculiar effects such as Klein tunneling and the ``Zitterbewegung'' (German for “trembling motion”) of electrons. From the time it was first identified by Erwin Schr\"odinger, Zitterbewegung (ZB) has been considered a key to understanding relativistic quantum mechanics. However, observing the original ZB of electrons is too difficult, and instead various emulations using entity models have been proposed, producing several successes.
Expectations are high regarding charge transports in semiconductors and graphene; however, very few reports have appeared on them.
Here, we report that ZB has a surprisingly large effect on charge transports when we play ``flat pinball'' with such trembling electrons in a semiconductor nanostructure.
The stage here is a narrow strip of InAs two-dimensional electron gas with a strong Rashba spin-orbit coupling. Six quantum point contacts (QPCs) are attached to the strip as pinball pockets. The ZB appeared as a large reproducible conductance fluctuation versus in-plane magnetic fields in the transport between two QPCs. Numerical simulations successfully reproduced our experimental observations, certifying that ZB causes a new type of conductance fluctuation.
\end{abstract}
\vspace*{-12pt}

In 1930, Erwin Schr\"{o}dinger \cite{Schrodinger} found that a free particle described by the relativistic Dirac equation undergoes oscillatory motion with the light velocity $c$. This phenomenon, called Zitterbewegung (meaning ``trembling motion'' and abbreviated as ZB),
originates from the pure quantum nature of relativistic velocity, which does not commute, as an operator, with the Dirac Hamiltonian.
This means the velocity, despite long being a most familiar quantity to physicists, is not a good quantum number for a free relativistic particle.
Although interest in this remarkable nature was stimulated among researchers, 
its estimated amplitude $\hbar/m_{0}c \sim$ 386~fm, and the frequency $2m_0c^2/\hbar\sim 1.6\times 10^{21}$~rad./s for electrons,
keep it very far out of experimental reach\cite{Zawadzki}. 
Successful ZB emulations have been achieved in entity models, including 
those with a single trapped ion\cite{Gerritsma} and ultra-cold atoms in Bose-Einstein condensates\cite{LeBlanc, Qu}.
However, clear observation of ZB in more realistic emulations with electrons themselves in artificial vacuums, 
namely in solids\cite{Schliemann, Rusin, Brusheim},
has been a major open task.
In two-dimensional electron systems with Rashba-type spin-orbit coupling (SOC)\cite{Rashba}, ZB is predicted to appear as a meandering of charge density\cite{Schliemann} with relatively large amplitude of more than ten nanometers \cite{Brusheim}, which is still barely observable by utilizing nanoscale techniques and superfine structures 
(although a possible observation was reported in a mesoscopic device\cite{Benter}).

\begin{figure}[b]
\centering
\includegraphics[width=0.45\linewidth]{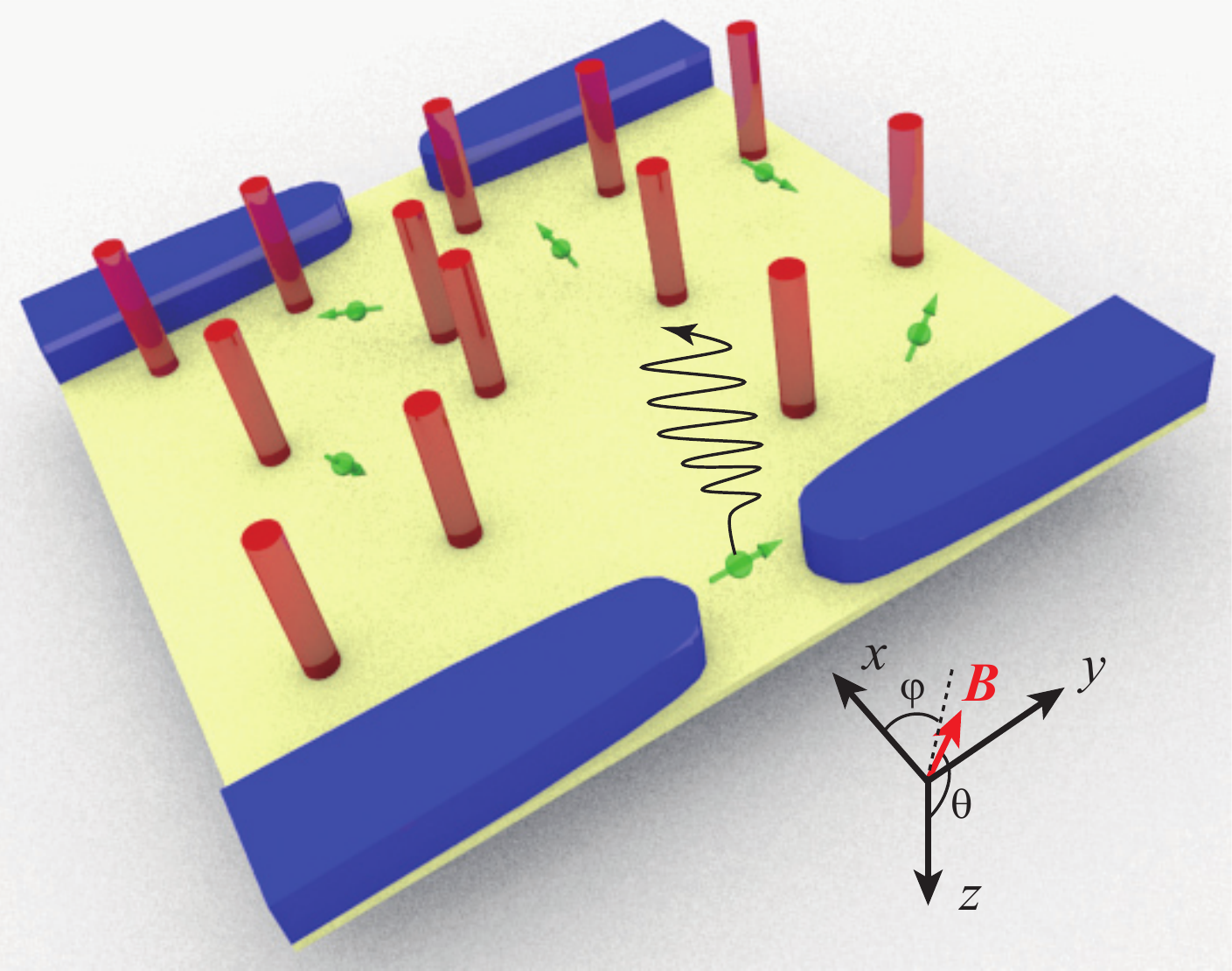}
\caption{{\bf Pinball on a flat table.} Electrons (green balls) are shot by QPCs (blue) into a 2DEG (yellow) with scattering by impurities (red cylinders).}
\label{fig0}
\end{figure}

Here, we report observation of ZB as reproducible conductance fluctuation (CF) versus magnetic fields in an InAs two-dimensional electron gas (2DEG) fabricated into an open billiard geometry, which has quantum point contacts (QPCs) as emitters and billiard pockets.
However, the experiment can be viewed as a pinball on a flat table (see Fig.\ref{fig0}) rather than as a billiard, because the table has a considerable number of fixed scatterers (impurities).
The ``flat pinball with trembling electrons'' model is also verified using a numerical simulation, which exhibits meandering of charge density and reproduces CF.


\section*{Results}
\subsection*{\bf ZB in Rashba model}
Let us briefly view how ZB appears in the Rashba model described by the Hamiltonian
\begin{equation}
H = \frac{\bm{p}^2}{2m^{*}}
+(\bm{a}+\bm{b})\cdot\bm{\sigma},\quad
\bm{a}\equiv\frac{\alpha}{\hbar}
\left(
\begin{array}{c}
p_y\\
-p_x
\end{array}
\right),\quad
\bm{b}\equiv
\frac{g^{*}}{2}\mu_{\rm B}\left( 
\begin{array}{c}
B_{x}\\ B_{y}
\end{array}
\right),
\quad
\bm{\sigma}\equiv
\left(
\begin{array}{c}
\sigma_{x}\\
\sigma_{y}
\end{array}
\right),
\label{eq_rashba_hamiltonian}
\end{equation}
where $\bm{p}$ is the electron momentum, $m^*$ the effective mass, $\alpha$ the Rashba SOC parameter, $g^*$ the effective Land\'e g-factor, $\mu_{\rm B}$ the Bohr magneton, 
$\bm{B}\equiv B_x\hat{x}+B_y\hat{y}$ (with $\hat{x}$, $\hat{y}$ as the unit coordinate vectors)
the in-plane magnetic flux density, and $\sigma_{x,y}$ the Pauli matrices.
As an initial condition ($t=0$), we consider an electron with a spin state of $z$-up in a Gaussian wavepacket at an origin
with central wave vector $\bm{k}=k_0\hat{k}_x$ (vector along $k_x$ with size $k_0$).
After a time evolution of $t$, the expectation value of the $y$-coordinate for finite $|\bm{a}+\bm{b}|$
is\cite{Biswas},
\begin{equation}
\Braket{y(t)} = \frac{\alpha}{2}\frac{(\bm{a}+\bm{b})_y}{[\hbar \omega(k_0)]^2}\{1-\cos[\omega (k_{0})t]\},
\label{eq_zb_y_1}
\end{equation}
where
\begin{equation}
\hbar\omega(k_0) = |\bm{a}+\bm{b}|,
\label{eq_precession_freq}
\end{equation}
while the spin portion of the wavefunction is written as $(\cos [\omega(k_0)t],\sin[\omega(k_0)t])$.
In the limit $\alpha\rightarrow 0$, the amplitude of the ZB in Eq.(\ref{eq_zb_y_1}) vanishes
while it approaches $k_0^{-1}$ for $\alpha\rightarrow\infty$.

\begin{figure}[b]
\centering
\includegraphics[width=0.9\linewidth]{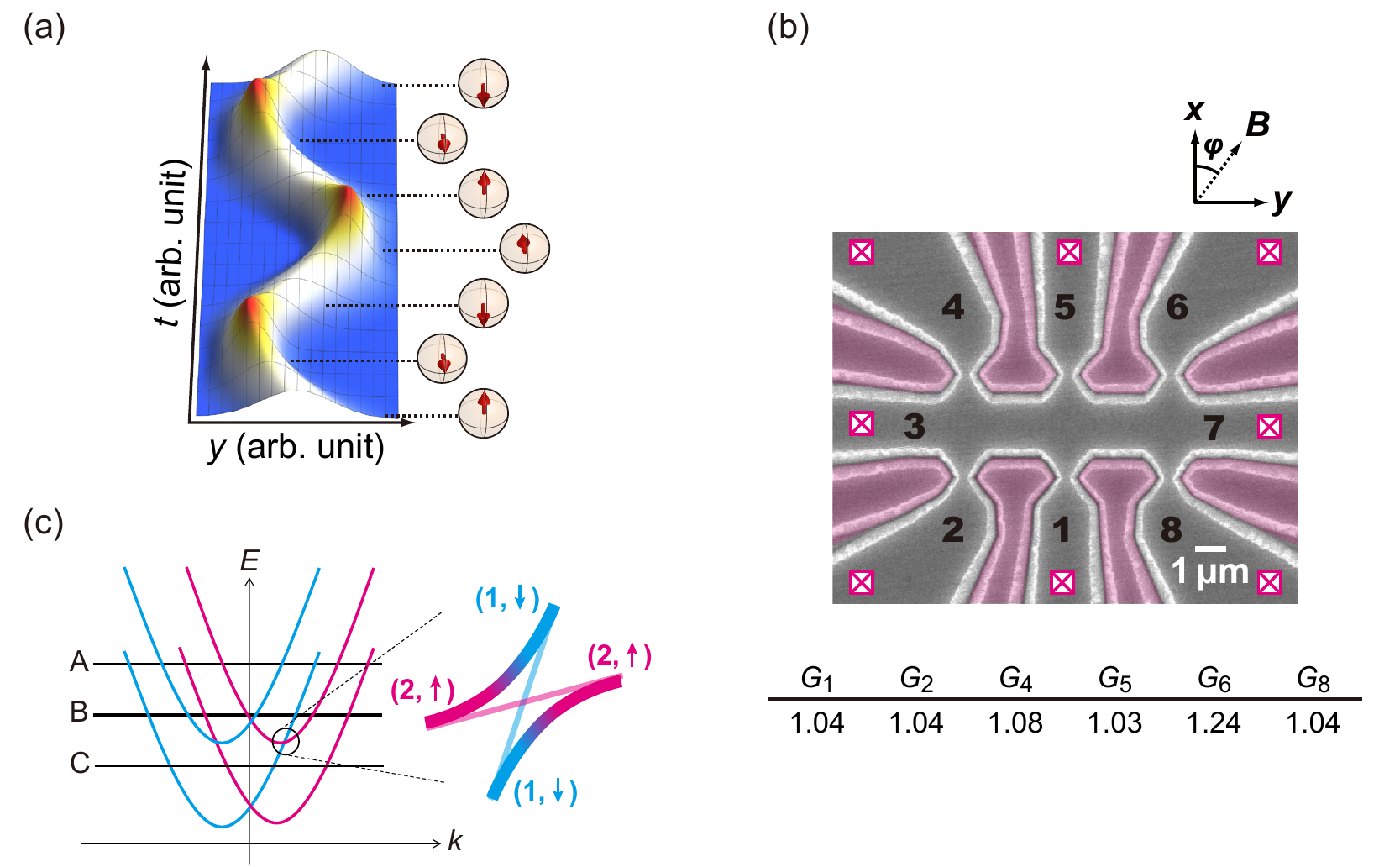}
\caption{{\bf Experimental setup.} (a) illustrates a ZB in the Rashba model.
An electron wavepacket, with a spin parallel to the $z$-axis and an averaged momentum along $x$ at time $t=0$, oscillates
along the $y$-axis in synchronization with the spin rotation in the $z$-$x$ plane.
(b) illustrates a scanning electron micrograph of the sample with the terminal numbers superposed. 
Pink false-colored areas act as gates isolated by trenches to define QPCs.
Below the micrograph, the conductances of QPCs are listed in the unit of $G_{\mathrm{q}}$ ($T=55~\mathrm{mK}$). 
(c) illustrates dispersions for one-dimensional bands in a QPC with a Rashba SOI.
The red and blue curves show up and down spins, respectively.
The extension of a crossing point reveals a small avoided crossing, which causes a spin rotation (and hence spin polarization) during an adiabatic transmission through the QPC.
A, C, and B indicate the positions of Fermi energy during the traversal — at the outlet, at the middle, and at a point between them,
respectively\cite{eto}.
}
\label{fig1}
\end{figure}

Figure \ref{fig1} (a) shows a schematic trace of the wavepacket oscillating along $y$ according to Eq.(\ref{eq_zb_y_1})
with spin directions.
As is easily understood, in a Rashba system,
ZB appears as a meandering electron orbit with a spin precession 
around the vector sum of $\bm{B}$ and an effective magnetic field originating from SOC
\begin{equation}
\bm{B}_{\rm eff} = \frac{2}{g^{*}\mu_{\rm B}}\bm{a}= \frac{2\alpha}{g^{*}\mu_{\rm B}}
\left(
\begin{array}{c}
k_0\\
0
\end{array}
\right),
\label{eq_effective_field}
\end{equation}
which is perpendicular to the momentum.

In the above formulas, the difficulties in the detection are apparent
in that 1) the ZB amplitude is the Fermi wavelength at most, and 2) to attain this amplitude the ZB frequency in Eq.(\ref{eq_precession_freq}) should be very high — above 1~THz.
In the following, we consider solving these problems in the introduction of the experimental setup.

\subsection*{\bf Experimental setup}
To create a Rashba system, a 2DEG in a pseudomorphic InAs quantum well was grown with molecular beam epitaxy.
The layered structure has asymmetric (In,Ga)As/InAs interfaces in the well region, whose structure is commonly adopted for
strong Rashba SOI\cite{Matsuda}.
The electron mobility $\mu$ = 6.6$\times10^4$~cm$^2$/Vs and the sheet carrier concentration $n = 1.1 \times 10^{12}~{\rm/cm^2}$
were obtained from the longitudinal and Hall resistances.
They produced the Fermi wavenumber $k_{\rm F}$= 2.6$\times10^8$~m$^{-1}$, 
which corresponds to wavelength $\lambda_{\rm F}=$2.4$\times10^{-8}$~m.
The effects of quantum confinement, wavefunction penetration into (In,Ga)As, and strain is renormalized into a shift in the in-plane effective mass in InAs from bulk value 0.023$m_0$ to
0.025$m_0$\cite{Yang}.
Then, the Fermi velocity $v_{\rm F}=1.2\times10^6$~m/s and the Fermi energy $E_{\rm F}=106$~meV,
$g^*=$8.6\cite{Yang2} were obtained.
The Rashba parameter $\alpha = 3.6\times 10^{-11}~{\rm eV m}$ 
was obtained from the amplitude modulation of the Shubnikov-de Haas (SdH) conductance oscillation\cite{nitta}.
These parameters give the effective magnetic field strength (\ref{eq_effective_field}) as 38~T, hence the ZB frequency
4.6~THz.

The high frequency problem can be solved by fixing the ZB in a steady state by using spin-polarized electrons.
This can be accomplished with a quantum point contact (QPC) on the conductance plateau of 
quantum conductance $G_{\rm q}\equiv 2e^2/h$, which was predicted theoretically\cite{eto} and confirmed experimentally\cite{kim}.
Figure \ref{fig1} (c) illustrates how the mechanism works to polarize spins of electrons passing through a QPC with Rashba SOI
on the conductance plateau of $G_{\rm q}$.
On the top of the effective potential in the QPC, the Fermi energy places at the lowest branch in the discrete dispersion relations. When an electron passes through the QPC adiabatically, its energy shifts as A $\rightarrow$ B $\rightarrow$ C $\rightarrow$ B' $\rightarrow$
A'. During the transition from C to B', the avoided crossing between (0$\downarrow$) and (1$\uparrow$) flips up the electron spin.
Owing to the ``one-way spin rotation,'' the spin polarization on plateau $G_{\rm q}$ reaches 0.7.
Here, the spin separation of the dispersion branch comes from the $\alpha p_x\sigma_y/\hbar$ term, and the avoided crossing comes from
$-\alpha p_x\sigma_y/\hbar$ in (\ref{eq_rashba_hamiltonian}).
The Rashba SOI is thus necessary for the spin polarization.

Figure \ref{fig1}(b) shows a scanning electron micrograph from the sample, on which the numbers of the terminals are indicated.
A right-handed spatial coordinate with a $z$-axis perpendicular to the 2DEG plane is taken as illustrated.
The structure was fabricated with electron beam lithography and dry etching of trenches.
Such trench-gate techniques are commonly used in InAs-based heterostructures\cite{kohda}, to which the Schottky gate technique is hardly applicable.
The structure consists of a main conducting strip (terminal 3-7) with a width of 2~$\mu$m and six QPCs opened on it.
A problem here is that the gates (pink false-colored regions) are common with the neighboring ones and individual control of QPC conductance is difficult. Fortunately, after several cut-and-tries, we succeeded in tuning the QPC conductances at around 1.0$G_{\rm q}$ with all the gates grounded (except QPC-6), as shown in Fig.\ref{fig1}(b).
The fact that such coarse tuning was successful indicates that the QPCs are on the 1.0$G_{\rm q}$ plateau, where the conductance is
less sensitive to the gate voltages than in other regions.

Then, in a conduction through any two of the QPCs, we emit spin-polarized electrons into the main strip region in one QPC and 
collect them through the other one.
Owing to the electron-hole symmetry, this holds even for conduction paths with QPC-6, in which the polarization may be insufficient.
However, the effective opening of the QPCs is around 500~nm, which is still much wider than the maximum amplitude of ZB.
Here, we attempt to solve the problem by using the scatterers in the strip as amplifiers of the meandering motions.
Because the width of the present mean free path $l_0\sim$1~$\mu$m is about half that of the main strip, 
the emitted electrons should experience a few scatterings before reaching a pocket.
If we approximate such a scattering with a classical scattering for a hard cylindrical wall with radius $R$, the scattering angle $\gamma$ for
impact factor $b_{\rm i}$ is $\gamma =2\cos^{-1}(b_{\rm i}/R)$.
When $b_{\rm i}$ oscillates against in-plane magnetic field $\bm{B}$ through a ZB with amplitude $\varDelta b_{\rm i}$,
the oscillation is amplified to $l_0\varDelta b{\rm i}/R$.
The potential range $R$ is roughly estimated to be an effective Bohr radius of 34~nm in InAs, which is even smaller than $k_{\rm F}^{-1}$.
Hence, the amplified ZB can reach the order of $l_0\sim 1$~$\mu$m, which can be resolved with the present QPCs.

The above rough classical sketch of scattering should be largely corrected in quantum mechanics, in that the spread of wave packets
weakens the amplification. As in the spin-polarization mechanism illustrated in Fig.\ref{fig1} (c), the wave packets emitted from a QPC should cover
at least $2k_{\rm F}^{-1}$.
However, as shown experimentally in scanning gate microscopy\cite{Topinka}, the wave packets emitted from a QPC
travel for surprisingly long distances without smooth spreading, owing to a kind of focusing effect.
Therefore, we can expect that the above classical model can be half-quantitatively applicable.
The experimental setup here can be viewed as a game of pinball on a flat table with six pockets.
The ball (wave packet) size may be larger than the pins (scattering centers), although minor wobbling in the orbits (the ZB)
is amplified by the scattering.

\subsection*{\bf Conductance fluctuations}
\begin{figure}[!]
\centering
\includegraphics[width=0.9\linewidth]{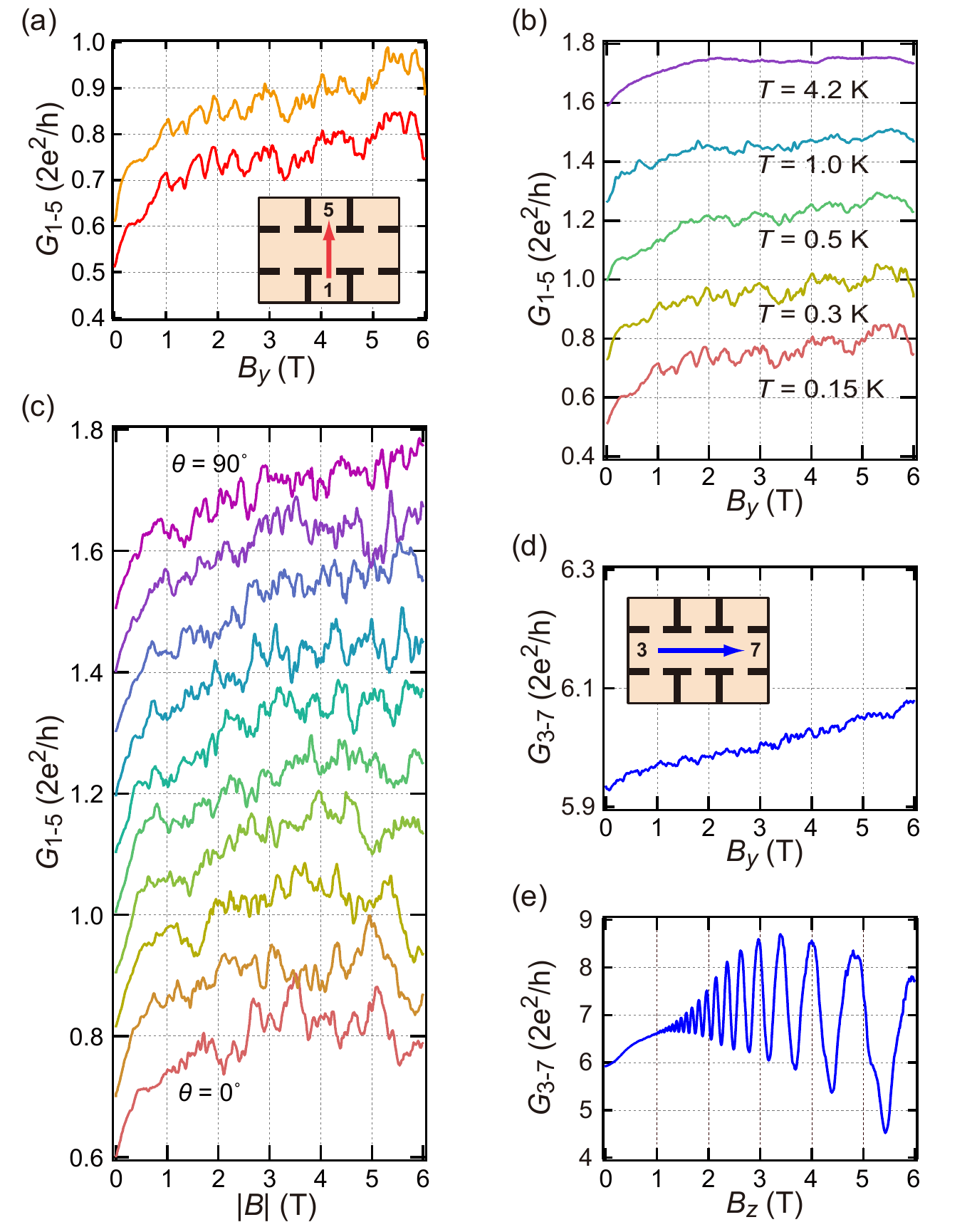}
\caption{{\bf Reproducible conductance fluctuation against a magnetic field.}
(a) $G_{1-5}$ is shown as a function of $B_{y}$ for two individual magnetic sweeps with offset = 0.1$G_{\rm q}$.
The inset shows the terminal configuration.
(b) $G_{1-5}$ for different temperatures with offset = 0.2$G_{\rm q}$.
In (c), $G_{1-5}$ at 70~mK is plotted versus the field amplitude $|B|$.
The direction of the magnetic field (the zenith angle $\theta$) is rotated from $\theta =0$ to 90$^\circ$ 
at azimuth angle $\varphi=$90$^\circ$.
The steps in the angle and conductance offsets are $10^\circ$ and $0.2 G_{\rm q}$, respectively. 
In (d) and (e), $G_{3-7}$ (the configuration shown in the inset) at 70~mK is plotted as a function of $B_{y}$ and $B_{z}$, respectively.}
\label{fig2}
\end{figure}

Henceforth, we use the notation $G_{i\mathchar`-j}$ for the two-wire conductance between terminals $i$ and $j$.
Figure \ref{fig2} (b) shows temperature variation of the two-wire conductance $G_{1\mathchar`-5}$ as a function of 
the in-plane magnetic field along $y$-axis $B_y$.
When the temperature was decreased, some aperiodic conductance oscillation versus $B_y$ became visible and the amplitude increased.
The oscillation is reproducible, that is, almost the same pattern appears for two independent field sweeps at the same temperature, as demonstrated in Fig. \ref{fig2} (a). 
Resemblances are also appreciable between the patterns for different temperatures.
Such conductance fluctuations (CFs), that is, the aperiodic response to an external magnetic field, were observed for an arbitrary combination of
two electrodes for QPC connections.

The result indicates that ZB is modulated by an external magnetic field and appears as CFs in the pinball model.
However, these remarkable features of CFs
are also reminiscent of so-called universal conductance fluctuation (UCF)\cite{Lee}.
We should thus check for the possibility of UCF.
Because the applied field is in-plane, we can eliminate the quantum interference effect through random paths tuned via the Aharonov-Bohm (AB) phase.
However, there still remains the possibility of interference in the spin portion of the wavefunction, which is modified by a spin precession tuned via the in-plane magnetic field (spin-UCF) \cite{Scheid}; for the elimination, we need the following two experiments.

The first is the rotation of the field direction from in-plane to perpendicular-to-plane (i.e., rotating the elevation angle from 0 to $\pi$/2).
The result is shown in Fig.\ref{fig2} (c). 
There is no significant variation in the amplitude and frequency distribution resulting from the pattern changes involving the angle.
This is hardly conceivable for interference type CF, because if there is such a network of spatial interference and the perpendicular component of the field is increased, the AB phase modulation participates in the interference modulation, which should bring some qualitative and systematic changes in the CF that stem from the geometric nature of the AB phase.
The second is the magnetic response of $G_{3\mathchar`-7}$, {\it i.e.}, a transport without QPCs (and hence without spin polarization).
Figures \ref{fig2} (d, e) display the result, which exhibits a dramatic change from Fig. \ref{fig2} (c), in that
in $G_{3\mathchar`-7}$ almost no fluctuation is observable in the in-plane field and ordinary SdH oscillation appearing in the perpendicular field (although 
the SOI modulation is reduced, probably by loose confinement into the strip).
This also indicates that the CFs are not from an interference because if such an interference exists, it inevitably appears in $G_{3\mathchar`-7}$, the path of which goes through the strip.

The most plausible origin of the CFs is now the ZB modulation of scattering in the pinball system; thus,
we further check the nature of CFs.
The pinball model has two temperature factors; one is the spin polarization at QPCs and the other is
the strength of the effective field $B_{\rm eff}$. 
The latter is related to Rashba splitting, which is approximately 6~K in the present case.
The CF is visible at order-of-magnitude lower temperatures.
Hence, the temperature dependence observed in Fig.\ref{fig2} (b) should come from spin polarization, 
which depends on the energy diagram inside the QPCs.
Derivation of the explicit temperature dependence is difficult, although the order estimation is reasonable
because the effective $E_{\rm F}$ is lower by an order of magnitude inside the QPCs.

\subsection*{\bf Azimuth angle dependence}
\begin{figure}[b]
\centering
\includegraphics[width=0.85\linewidth]{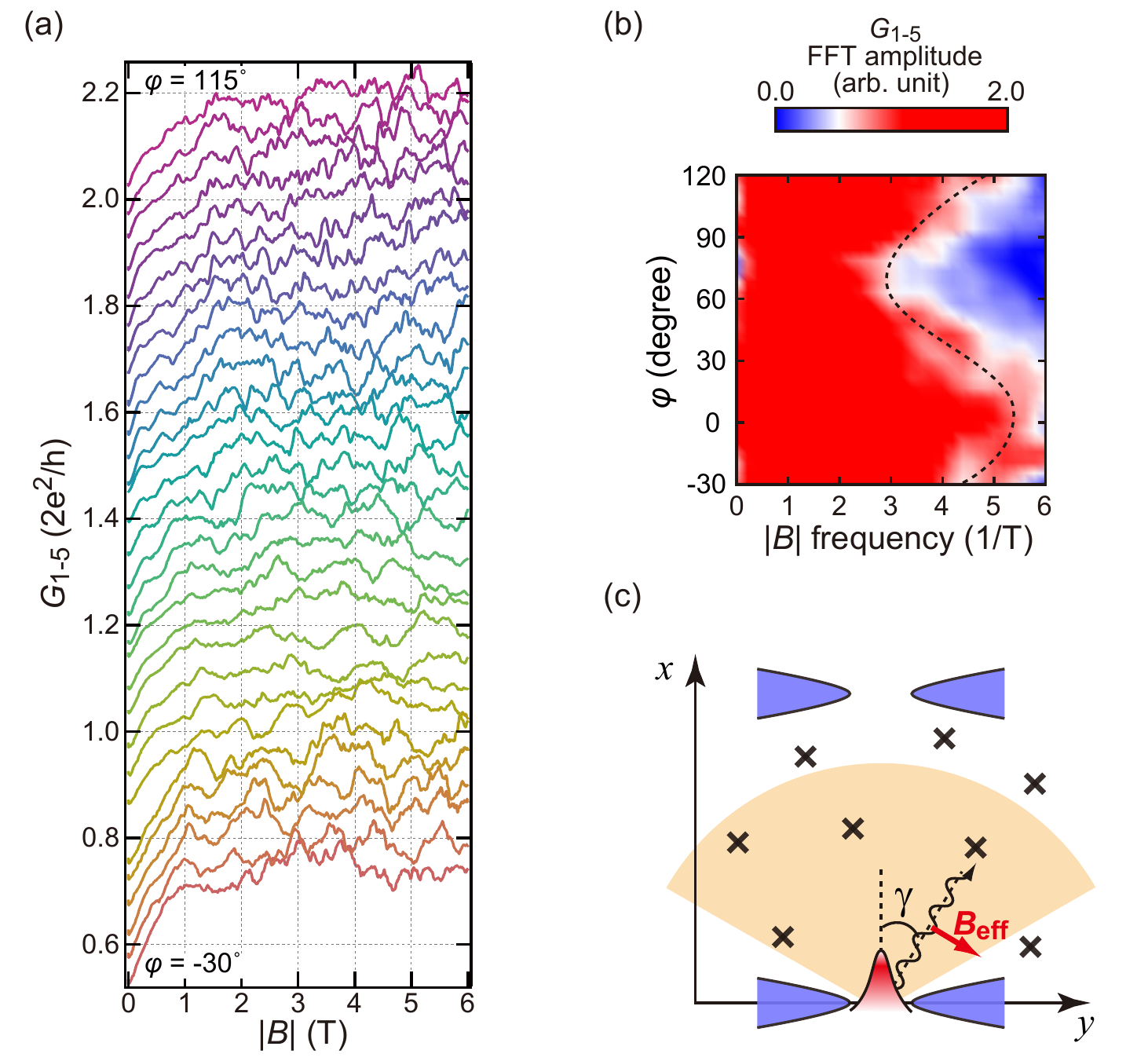}
\caption{{\bf Effect of azimuth angle in the magnetic field on the conductance fluctuation.}
(a) Conductance fluctuation in $G_{1-5}$ at 150~mK against the magnetic field amplitude $|B|$
for different azimuth angles $\varphi$ from $-30^\circ$ (bottom) to $115^\circ$ (top). 
The step in $\varphi$ and the offset in $G_{1-5}$ are $5^\circ$ and $0.05 G_{\rm q}$, respectively. 
(b) The FFT amplitude spectrum of the conductance fluctuation in a $|B|$-range from 1.5 to 6~T is plotted in color for $\varphi$.
The broken line represents a smoothed $\varphi$-tendency in Eq.\eqref{eq_angular_dep}.
(c) A schematic illustration of an electron's path. The orange area is the ``fan" where electrons can fly.
}
\label{fig3}
\end{figure}

Figure \ref{fig3} (a) shows the variation of CF on the azimuth angle $\varphi$ for an in-plane magnetic field.
The fluctuation pattern changes with $\varphi$  according to its resemblance to neighboring patterns.
A frequency amplitude spectra obtained with a fast Fourier transform (FFT) is shown in Fig.\ref{fig3} (b); it indicates some
systematic angular dependence in the width of spectra.

Let us check the correspondence between the pinball game model and the experimental results.
Because the spin polarization is along the $y$-axis directly near the emitter QPC\cite{eto}, the electrons with momentum parallel to the $x$-axis do not tremble because $\bm{B}_{\rm eff}$ is perpendicular to the momentum from Eq.(\ref{eq_rashba_hamiltonian}), which is parallel to the spins.
Now we consider a path with an oblique angle $\gamma$ to the $x$-axis (Fig.\ref{fig3}(c)).
The length of $\bm{B}_{\rm eff}+\bm{B}$ is approximated as $B_{\rm eff}+B\cos(\gamma+\phi)$,
where $B$ is the external field strength and $\phi\equiv \pi/2-\varphi$ for $|\bm{B}_{\rm eff}|\gg |\bm{B}|$.
The ``wavelength" of ZB is estimated as $\lambda_{\rm ZB}\equiv \pi\hbar^2/\alpha m^* \approx$ 270nm 
and the maximum modification of $\lambda_{\rm ZB}$ with an external field for $B$ =6~T and $\gamma+\phi =0$ is 34~nm.
This means only a single cycle of meander changes in the traversal of 2~$\mu$m for $B=$6~T.

However, this does not mean maximum frequency 1/6~T$^{-1}$ is expected in the CL.
Because a channel opening of a scattered orbit is for a very narrow range of magnetic fields, the conductance for the opening appears as a narrow
peak, which contains high-frequencies magnetic responses.
As illustrated in Fig.\ref{fig3}(c), if there are $N$ scatterers inside a ``fan" of initial momentums, there should appear $N$ to $2N$ such conductance peaks
corresponding to channels with a single scattering within 6~T, forming a fluctuation pattern. 
The density and sharpness of conductance peaks increases for channels with more scatterings, while the peak height decreases.
In the FFT power spectra, such gradual tailing to high frequencies is actually observed.
In the conduction from terminal 5 to 1, the paths should distribute around $\gamma=0$.
Thus, we further approximate $B\cos(\gamma+\phi)\sim -B\cos\gamma\sin\varphi$.
Further, the $\varphi$-dependence of the power spectral width $W_{\rm p}$ is roughly written as
\begin{equation}
W_{\rm p}(\varphi)\propto \frac{l_0}{\lambda_{\rm ZB}^2}\left|\frac{\varDelta \lambda_{\rm ZB}}{\varDelta B}\right|=l_0\frac{\pi g^*\mu_{\rm B}}{v_{\rm F}}|\sin\varphi|,
\label{eq_angular_dep}
\end{equation}
where $l_0$ is the averaged ZB path length.
Although the approximation is very coarse, Eq.\eqref{eq_angular_dep} suggests that $W_{\rm p}$ takes a maximum around 90$^\circ$ and a minimum around 0$^\circ$, which is clearly recognizable in Fig.\ref{fig3} (b).

\begin{figure}[b]
\centering
\includegraphics[width=0.9\linewidth]{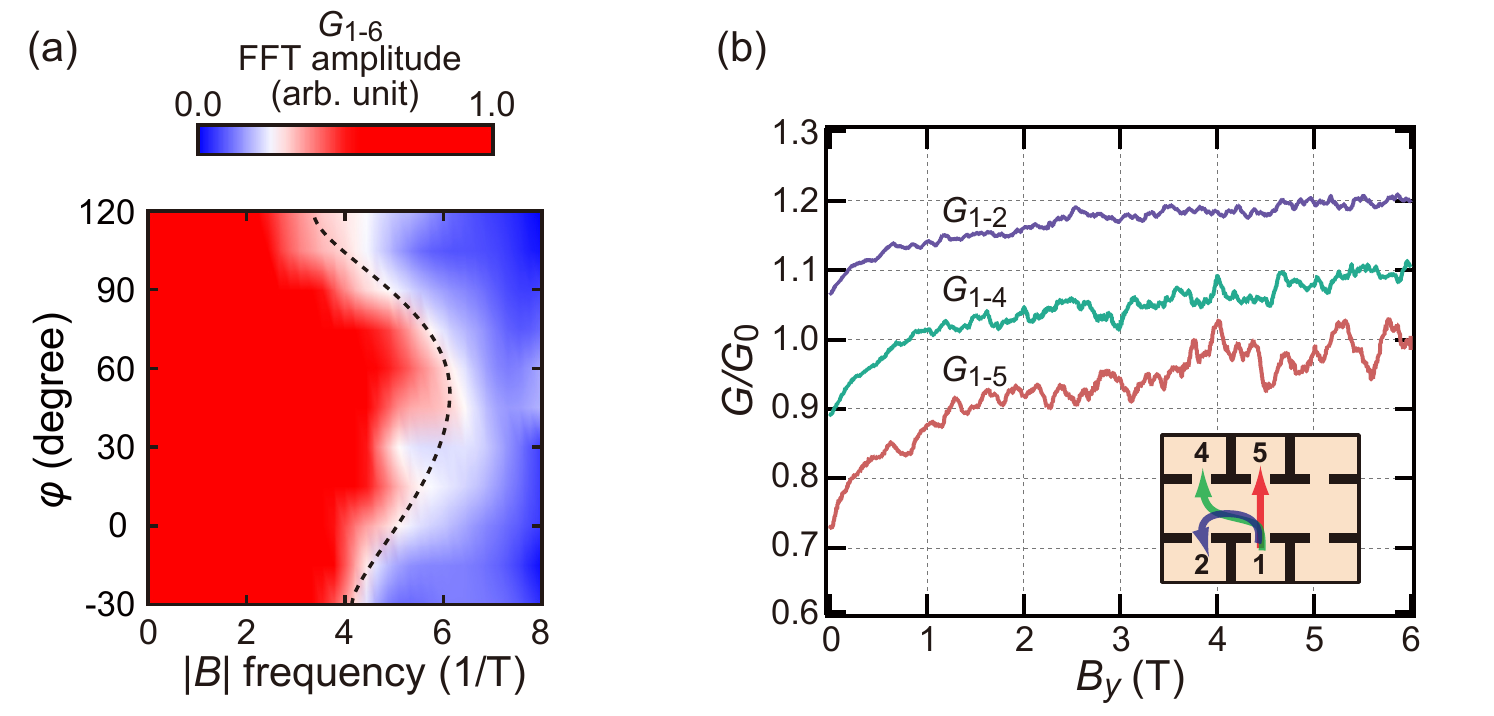}
\caption{{\bf Effect of probe configuration.}
(a) The same FFT map as Fig.\ref{fig3} (b) for $G_{1-6}$ at $T=150~\mathrm{mK}$. 
(b) $G_{1-2}, G_{1-4}$, and $G_{1-5}$ versus $B_{y}$, of which the configurations are illustrated in the inset.
The data are normalized by $G_{0} = G(B_{y}=6~\mathrm{T})$ and displayed with offset $0.1 G_{\mathrm{q}}$.}
\label{fig5}
\end{figure}

An interesting test is to change the combination of electrodes.
Figure \ref{fig5} (a) shows results of the same experiment as Fig.\ref{fig3} (b) but for the conductance between electrodes 5 and 2.
Now, the center of path distribution is around $\gamma=\pi/4$, and $\varphi$-dependence in Eq.\eqref{eq_angular_dep}should change to $|\sin(\pi/4+\varphi)|$ as is indeed observed in Fig.\ref{fig5} (a).
Figure \ref{fig5} (c) shows the CLs for three different combinations of electrodes under an external magnetic field with $\varphi=\pi/2$.
The difference between $G_{1\mathchar`-5}$ and $G_{1\mathchar`-4}$ can be explained by a $\pi/4$ shift in $\varphi$-dependence.
More interesting is $G_{1\mathchar`-2}$, in which backscatterings are inevitable; hence, paths with multiple scatterings survive,  resulting in reduced amplitude fluctuation and higher frequencies.
In $G_{1\mathchar`-3}$, the QPC path selection mechanism QPC does not work, although coupling with transverse modes should be
required. The ZB-induced oscillation in the coupling causes residual conductance fluctuation with much smaller amplitude.

\subsection*{\bf Tight binding calculation}
Thus far, we have checked the ZB in a simple pinball model from various viewpoints.
Here, we show that ZB and impurity scattering  causes significant fluctuation in the conductance between QPCs in the Rashba model
with numerical calculations.

\begin{figure}[b]
\begin{center}
\includegraphics[width=0.8\linewidth]{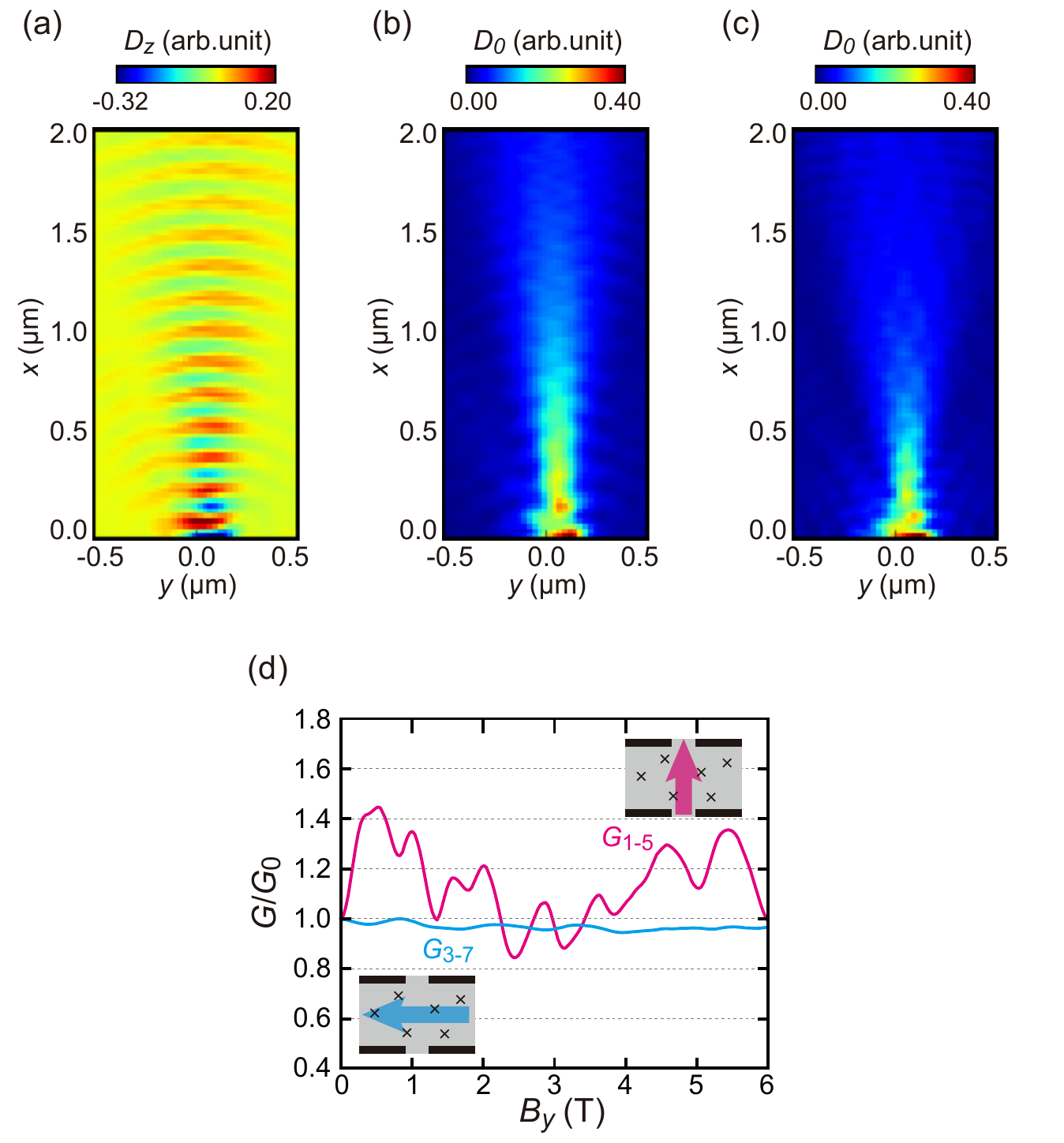}
\end{center}
\caption{{\bf Numerical simulation with ZB reproduces the conductance fluctuation.}
(a) and (b) are calculated spatial distributions of the $z$-component of spin ($D_{z}$) 
and the probability amplitude ($D_{0}$), respectively, for a two-dimensional electron propagating from the emitter QPC at the origin under a zero-magnetic field.
The boundary condition of the spin at the QPC is taken to be polarized along $x$, and no scattering center is introduced.
(c) The same plot of $D_{0}$ as (b), but with $B_{y} = -3.5~{\rm T}$. 
(d) Calculated $G_{1-5}$ (with polarization) and $G_{3-7}$ (without polarization) are shown as a function of $B_{y}$.
Random scattering centers are introduced.
The data are normalized by $G_{0} = G(B = 0)$. The probe configurations are illustrated in the insets.}
\label{fig6}
\end{figure}

Figure \ref{fig6}(a-c) shows the spatial distribution of spin-polarized electrons in a steady flow from a narrow constriction
into a clean two-dimensional region.
Clear meandering is observed, confirming the existence of ZB.
Figure \ref{fig6}(d) displays conductance by electrons with and without spin-polarization 
between two constrictions through a strip region with scatterers.
The conductance shows a large aperiodic response to an in-plane magnetic field for spin-polarized electrons, while
such fluctuation is much smaller for unpolarized electrons.
Note that we take an absorptive boundary condition, which breaks the unitarity. 
This is to kill unrealistic conductance fluctuation due to interference with infinite coherence.
The residual conductance fluctuation for spin-unpolarized electrons should come from such interference on paths
without bounce to the boundaries.
The result indicates that, in a Rashba system with scatterers, ZB truly causes conductance fluctuation,
supporting some reality of the pinball model.

\section*{Methods}
\subsection*{Two-dimensional electron system in an InAs quantum well}
The substrate was an InAs quantum well, grown on a (001) InP using the molecular beam epitaxy technique. It consists of, from the bottom, InAlAs and InGaAs (700 nm) as buffer layers, InAs (4 nm), InGaAs (4 nm), InAlAs (10 nm), n-InAlAs (40 nm), InAlAs (5 nm), and InGaAs (2 nm). QPCs were fabricated by electron-beam lithography and Ar dry-etching with the depth of 300 nm. Contacts were fabricated by AuGe deposition (100 nm) and subsequent annealing (280 degree, 5 min).

The specimen was cooled to 70~mK in a dilution fridge with a superconducting solenoid for applying the magnetic field.
A conventional lock-in technique was used for measuring two-wire conductances with frequencies lower than 1 kHz.

\subsection*{Setups for numerical calculations}
For numerical calculations we employed the ``Kwant'' package\cite{Kwant}, which is based on a tight-binding approximation. The distance between two QPCs $L=1.8~{\rm \mu m}$, QPC width $w = 400~{\rm nm}$, and Rashba strength $\alpha = 3.6\times 10^{-11}~{\rm eV m}$ were adopted to simulate the experiment. The effective mass $m^{*} = 0.025 m_{0}$ (where $m_{0}$ is the electron mass in vacuum) and g-factor $g^{*} = +8.6$ were adopted from previous studies \cite{Yang}. 
$E_{\rm F}$ was tuned to give the QPC conductance $G_{\rm QPC} = 1.0 G_{\rm q}$. The calculated area was $3.6~\mu{\rm m}\times 3.0~\mu{\rm m}$, and was surrounded by absorption walls (except for the QPC passes) and simplified to a square lattice with the lattice constant $a=25~{\rm nm}$. Impurities were introduced by adding randomly-distributed values to onsite energy. The typical amplitudes of the impurities were $0.2~{\rm eV}$.

\section*{Acknowledgment}

This work was supported by a Grant-in-Aid for Scientific Research on Innovative Areas, ``Nano Spin Conversion Science" (Grant No. 26103003), also by Grant No. 25247051 and by Special Coordination Funds for Promoting Science and Technology. 
Iwasaki was also supported by the Japan Society for the Promotion of Science through Program for Leading Graduate Schools (MERIT).

\section*{Author contributions}
Y. I. fabricated the sample and performed measurements.
Y. H grew and provided the substrate.
Y. I, T. N and S. K discussed the results and derived the theoretical model.
Y. I and S. K wrote the manuscript.

\end{document}